\documentclass{aa}

\usepackage{astro}
\usepackage[final]{graphicx}
\usepackage{txfonts}
\usepackage{natbib}
\usepackage[LGR,T1]{fontenc}
\usepackage[latin1]{inputenc}

\long\def\correct#1{%
   \ifmmode
      \mathbf{#1}%
   \else
      \mathversion{bold}%
      \textbf{#1}%
      \mathversion{normal}%
   \fi%
}
\def\degree{\ensuremath{^\text{o}}}
\begin{document}


\title{The vertical structure of T~Tauri accretion discs}
\subtitle{IV. Self-irradiation of the disc in the FU~Orionis outburst phase}

\author{R. Lachaume\inst{1}}

\institute{%
   Max-Planck-Institut f\"ur Radioastronomie,
   Auf dem H\"ugel 69, D-53121 Bonn\\
   \email{lachaume@mpifr-bonn.mpg.de}
}

\date{Received; accepted}
   
\abstract{%
   I investigate the self-irradiation of intensively accreting circumstellar
   discs (backwarmed discs).  It is modelled using the two-layer disc approach
   by \citet{Lachaume03b} that includes heating by viscous dissipation and by
   an external source of radiation.  The disc is made of a surface layer
   directly heated by the viscous luminosity of the central parts of the disc,
   and of an interior heated by viscosity as well as by reprocessed radiation
   from the surface.  This model convincingly accounts for the infrared excess
   of some FU Orionis objects in the range 1--200\,{\micron} and supports the
   backwarmed disc hypothesis sometimes invoked to explain the mid- and
   far-infrared  excesses whose origins are still under debate. Detailed
   simulation of the vertical radiative transfert in the presence of
   backwarming is still needed to corroborate these results and
   spectroscopically constrain the properties of intensively accreting discs.
}

\maketitle 

   
 
\section{Introduction}

It is now widely accepted that circumstellar discs accompany the process of
star formation, all the more as such discs have already been imaged in the
millimetre \citep[ex. DG~Tau,][]{Dutrey94} and in the infrared \citep[ex.
HK~Tau,][]{Stapelfeldt98}.  Matter accreting from the disc onto the star is
supposed to build up solar-mass stars in the timescale of a \Myr; yet recent
studies of star forming regions indicate way too low accretion rates among
T~Tauri stars (TTS) ---in the range $10^{-10}$--$10^{-7}\,\Msun/\yr$, as shown
by \citet{Gullbring98} in the Taurus region and \citet{Robberto03} in the
Trapezium.  Part of the matter is supposed to be accreted in earlier phases of
high accretion but young stellar objects (YSOs) are then so embedded that their
disc cannot be optically observed \citep[see][for the stages of
evolution]{Andre94}.  The other part might also be accreted during brief,
periodic phases of intense accretion of otherwise quiescent TTS
\citep{Hartmann96}:  FU~Orionis objects (FUors), which feature accretion rates
up to $10^{-4}\,\Msun/\yr$ and which are seen to undergo an increase in
luminosity of more than 4 visual magnitudes in a few years
\citep{Herbig66,Herbig77}, may represent such a phase.

FUors are convincingly modelled by a self-heated viscous accretion disc (active
disc) that overwhelms the stellar light and the properties of FUor outbursts
are also well-studied \citep{Hartmann85,Bell95}.  Yet active disc models fail
to explain the mid- and far-infrared excess \citep[eg.][]{Simon88} as well as
weak silicate feature at $10\,\micron$ sometimes
present in emission \citep{Hanner98}; these properties seem more typical of an irradiated disc.
Alternative models have tackled this problem:  \citet{Lodato01} showed that the
self-gravity of the disc can trigger an instability that produces additional
warming and were able to reproduce the SED.  However, this model cannot account
for the silicate feature, which requires a temperature inversion at the surface
of the disc.  A more pleasing approach is the presence of a circumstellar
envelope proposed by \citet{Adams87}: on one (observational) hand,
\citet{Kenyon91} successfully fit the spectra of two FUors with such an
envelope.  On the other (theoretical) hand, the envelope could serve as a
reservoir of  infalling material, replenishing the disc between outbursts. On the
third hand, the envelope can account for a part of the extinction observed in
these objects. 

\citet{Kenyon91,Bell99} also proposed that the inner hot parts of the disc are
bright enough to heat up the outer ones (backwarming) and produce an irradiated
disc-like SED.  Their simulation uses a black-body disc model, ie.  without
vertical temperature profile, and convincingly explains the order of the excess
at 30--100\,\micron, yet it fails to reproduce the SED at 10--30\,{\micron} and
does not predict a silicate feature in emission.
\citet{Malbet91,Calvet91,Chiang97} showed that irradiation can produce a hot
disc surface radiating at shorter wavelengths and account for emission features
(CO bands, silicates).  Using their results, \citet{Lachaume03b} developed a
two-layer disc model in which irradiation by a central star and viscosity are
taken into account.  In this paper, I shall use this model with the disc itself
as a source of radiation and investigate the backwarming in FUors.

\section{The model}

\begin{figure}[t]
   \centering
   \includegraphics[width=0.75\hsize]{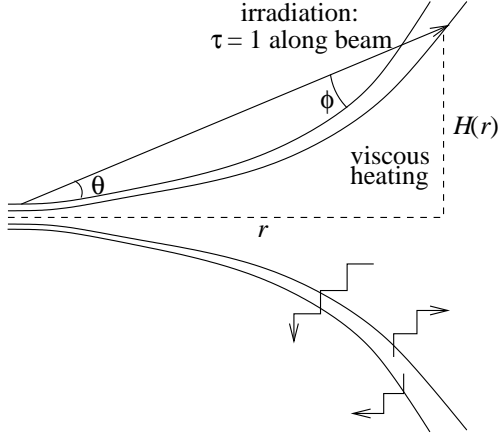}
   \caption{Geometry of a backwarmed disc. The incidence angle $\phi$ of
   the incoming radiation onto the surface and the angle $\theta$ at which
   the inner regions are seen from it depend on the profile of the surface
   $H(r)$.}
   \label{fig:scheme}
\end{figure}

I use the two-layer approach described in detail by \citet{Lachaume03b}:  the
surface of the disc is directly irradiated by the incoming radiation with an
optical thickness of unity in the visible along the slanted path of the incident
beam.  Its vertical optical thickness in the infrared, to which radiation is
mostly reprocessed, is then much smaller than one.  Therefore, it is at a
higher temperature than its effective temperature.  The interior of the disc is
both backwarmed by the surface and heated by viscous dissipation, modelled with
the $\alpha$ viscosity prescription by \citet{Shakura73}.  A scheme of the
model is presented in Fig.~\ref{fig:scheme}.

\subsection{Irradiation}

Since most of the flux originates from the first few solar radii of the disc,
the irradiation source is seen as a central, infinitely flat ring
from the the outer regions of the disc (1--100\,\AU).  The radial flux
distribution is, according to the standard disc model by \citet{Shakura73},
\begin{equation}
   F(r)   = \frac{3}{8} \left(1-\sqrt{\frac{r_0}{r}}\right) \frac{\Gravity\Mstar\Mdot}{\pi r^3},
\end{equation}
where $\Mstar$ is the mass of the star, $r_0 < \Rmin$ a constant, and
$\Rmin$ the inner radius of the disc.  The value of $r_0$ depends on
the torque exerted on the inner rim of the disc.  If the latter extends down to
the star, the standard model assumes that the former is linked to the
deceleration of the matter falling from the supersonic disc to the subsonic
equator of the star, and $r_0 \approx \Rstar$, the stellar radius. However, the
torque exerted on a disc with an inner gap remains unknown; it should depend on
the magnetic coupling between the star and the disc.  Without clear theoretical
or observational hints, I decided to assume $r_0 = \Rstar$, as in the standard
model.

The total luminosity of the disc amounts to
\begin{equation}
\begin{split}
   \Fdisc &= \int_{\Rmin}^{+\infty} 2\pi r F(r) \idiff r\\
          &= \left(3-2 u \right) u^2 \, \frac{\Gravity\Mstar\Mdot}{4\Rstar},
\end{split}
\end{equation}
where $u = \sqrt{\Rstar/\Rmin} < 1$. At a given location on the disc's 
surface, the heating flux, ignoring the shadowing of some
parts of the disc by the star, is:
\begin{equation}
   \Fheating = (2/3+\cos\theta) \sin\theta \sin\phi \, \frac\Fdisc{r^2 + H^2},
   \label{eq:Fheating}
\end{equation}
where $\theta$ is the angle at which the heating ring is seen, $\phi$
the incidence angle onto the surface of the disc, and $H$ the altitude
of the surface above the mid-plane.  In Eq.~\ref{eq:Fheating}
the first term models the ``limb-darkening'', while the second and
third ones are projection effects.  With an infinitely flat ring
\begin{align}
   \theta &= \Arctan \frac Hr,\\
   \phi   &= \Arctan \left( \deriv Hr \right) - \theta.
\end{align}

The characteristic temperature of the irradiation is needed to determine
the optical thickness of the surface layer ---it depends on the penetration
of the incoming radiation.  We use the flux-weighted average of the  
viscous effective temperature $\Tv(r)$:
\begin{equation}
\begin{split}
   \Tirr^4 &= \frac 1\Fdisc \int_{\Rmin}^{+\infty} 2\pi F(r) \Tv(r)^4 \idiff r\\
           &= \frac{45 - 80 u + 36 u^2}{3 - 2 u} u^6 \, \frac{\Gravity\Mstar\Mdot}{160\sigma\pi\Rstar^3}
\end{split}
\end{equation}

\subsection{Location of the surface}

\citet{Chiang01} provide a method to consistently find the location of
the irradiated surface in an irradiated two-layer disc.  However, the method
is more difficult to implement in models that also include viscous heating,
in which viscosity-dominated regions may present self-shadowing.  Instead,
we used a simpler approach proposed by \citet{Chiang97} and state that
irradiation-dominated regions meet
\begin{equation}
   H \propto h, \label{eq:H/h}
\end{equation}
where $h$ is the vertical pressure scale-height of the disc.  It appears that
such a prescription results in a surface profile $H(r)$ close to that
predicted by the self-consistent method in the irradiation-dominated regions
\citep{Lachaume03b}.  Though one expects it to fail in viscosity-dominated
regions, the impact on the disc structure is low, since these regions are by
definition not affected by the irradiation.

We now look for a self-similar solution $H/r = r^\gamma$.  Within the
approximation $H(r)/r \ll 1$, Eq.~\ref{eq:Fheating} reads
\begin{equation}
   \begin{split}
      \Fheating &\approx 2/3 \left(\frac{H}{r}\right) \left((\gamma-1) \frac{H}{r}\right) \frac1{r^2}\\
                &\propto h^2 r^{-4}. 
   \end{split}
   \label{eq:flaring:1}
\end{equation}
The pressure scale height is obtained from the vertical gravity field of the 
star:
\begin{equation}
   h \approx r^{3/2} \sqrt{\frac{\kB\Tc}{\Gravity\Mstar\mH}}
\end{equation}
where
\begin{equation}
   \Tc = (\Fheating/2)^{1/4}
\end{equation}
is the temperature of the interior of the disc in irradiation-dominated 
regions. The last two equations, together with $H \propto h$, yield
\begin{equation}
   H \propto r^{3/2} \Fheating^{1/8}. \label{eq:flaring:2}
\end{equation}
By subsituting Eq.~\ref{eq:flaring:1} into Eq.~\ref{eq:flaring:2} one
finally derives
\begin{equation}
   H/r \propto r^{1/3}.
\end{equation}
This determination does not provide a estimate for the proportionality
constant, which I leave as a free parameter in the model.

\begin{figure*}[t]
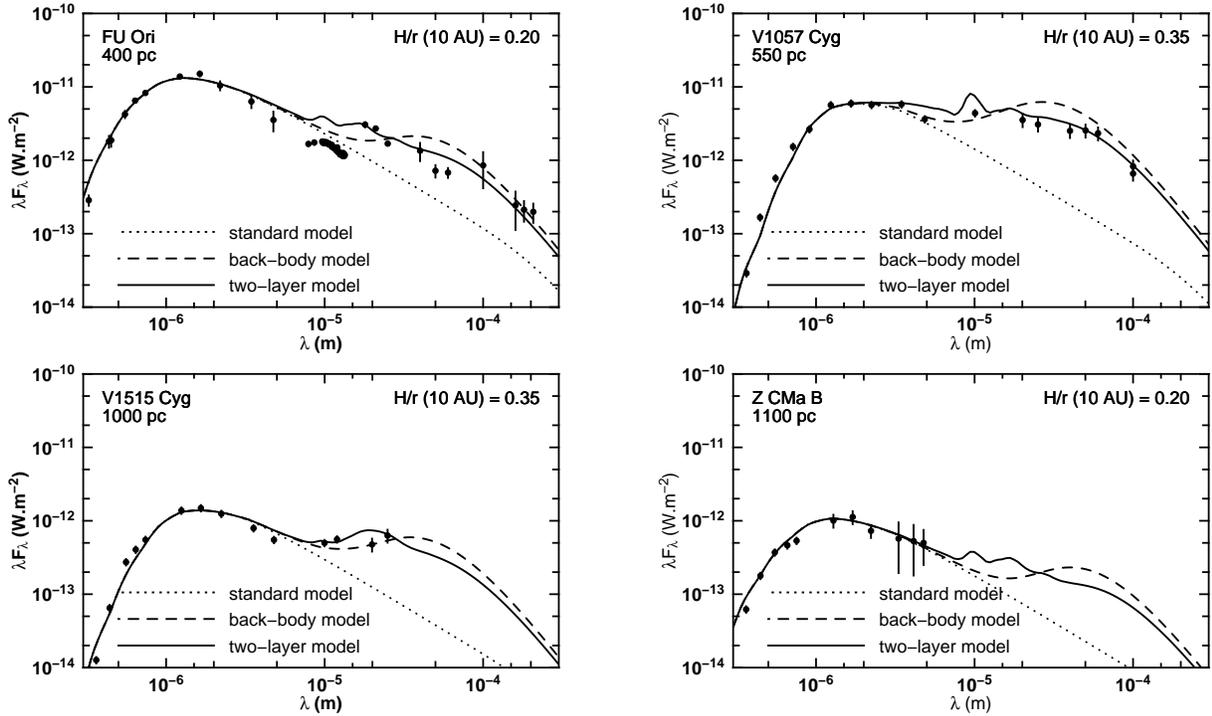

   \centering
   \includegraphics[width=0.45\hsize]{fig2a.ps}%
   \hskip 0.03\hsize
   \includegraphics[width=0.45\hsize]{fig2b.ps}%
   \\[-3ex]
   \includegraphics[width=0.45\hsize]{fig2c.ps}%
   \hskip 0.03\hsize
   \includegraphics[width=0.45\hsize]{fig2d.ps}%
   \caption{The SED of FUors.  
   Solid line: two-layer disc model including viscous
   heating and backwarming.  Dashed line: black-body disc model
   including viscous heating and backwarming.
   Dotted line:  disc model with viscous heating
   alone.}
   \label{fig:fuor-sed}
\end{figure*}

\section{Results \& Discussion}

\subsection{Strategy}

\begin{table}[t]
   \centering
   \def\nc{\rlap{$^\star$}}
   \caption{Fundamental parameters of FUor models: accretion rate, inner
      radius of the disc, flaring $H/r$, type of dust opacity $\kappa_\lambda$, and
      extinction $A_V$.
   }%
   \label{tab:param-fit}
   \begin{tabular}{lccccc}
      \hline\hline
                & $\Mstar\Mdot$   & \Rmin    & $H/r$        & $\kappa_\lambda$ & $A_V$\\
                & ($\Msun^2/\yr$) & (\Rsun)  & (at 10\,\AU) & (ref.)           &      \\
      \hline
      FU~Ori    & \sci{3.2}{-5}   & 4        &  0.20        & i                & 1.3\\
      V1057~Cyg & \sci{2.0}{-5}   & 2        &  0.35        & i                & 4.5\\
      V1515~Cyg & \sci{1.3}{-5}   & 5        &  0.35        & h                & 2.8\\
      Z~CMa B   & \sci{1.4}{-5}   & 2        &  0.20\nc     & i                & 1.8\\
      \hline
      \multicolumn{6}{l}{i: inhomogeneous aggregates \citep{Henning96}}\\
      \multicolumn{6}{l}{h: homogeneous aggregates \citep{Henning96}}\\
      \multicolumn{6}{l}{$^\star$: not constrained}\\
      \hline
   \end{tabular}
\end{table}

I apply this model to unembedded FUors, since more embedded objects cannot be
properly described without considering reprocessing of the disc emission by the
surrounded dusty material, which is beyond the scope of this paper.  As a
criterion for ``unembeddedness'' I choose a decreasing SED in the infrared,
which leads me to consider three confirmed FUors, \object{FU~Ori},
\object{V1057~Cyg}, and \object{V1515~Cyg}, and one FUor candidate,
\object{Z~CMa} \citep{Kenyon91}.  Their unembeddedness is confirmed by their
limited extinction $A_V \lesssim 4$\,mag \citep{Cohen79,Bell95}.  Photometry
data are taken from \citet{Kenyon91} for V1057~Cyg and V1515~Cyg, from
\citet{Thiebaut95} for Z~CMa, and from the \citet{Gezari99} catalogue for
FU~Ori.

The models have been computed using wave\-length-de\-pen\-dent grain opacities
for dust aggregates by \citet{Henning96}, needed for the determination of the
optical thickness of the surface layer and for the computation of the SED.  At
higher temperatures, I used grey gas opacities by \citet{Bell94}.
Model parameters \Rmin, $\Mstar\Mdot$, and {\Av} can be determined by
first fitting an active disc model to the SED in the range 0.3--3\,{\micron};
they do not depend on the strength of the mid-IR excess to be modelled.  The
constant $H/r\ (10\,\AU)$ is then adjusted to fit the backwarmed disc model to
the mid- and far-IR excess. $\Mstar\Mdot$ and $H/r\ (10\,\AU)$ are
unambiguouisly constrained (by the 1--3\,{\micron} flux and the
10--50\,{\micron} SED respectively), while $\Rmin$ and $\Av$ are derived from
the SED at $\lambda \lesssim 0.5\,\micron$.

In all models, the viscosity parameter is $\alpha = 0.1$ and the
albedo of the surface $\albedo = 0.2$. $\alpha$ has almost no influence on the
SED at $\lambda \lesssim 100\,\micron$ because these discs are optically thick,
and the presence of the albedo increases the visible and near-IR SED by a few
percent while decreasing the mid- and far-IR excess by up to 20\,\%.  Z~CMa
and FU~Ori, which exhibit observational hints of not beeing seen pole-on, are
assumed to be seen with $i = 40\degree$, other targets are modelled
as pole-on ($i = 0\degree$).  The inclination has little influence
for angles $i \lesssim 45\degree$ and mostly scales down the visible and
near-IR SED by a factor \correct{$\cos i$}.

\subsection{Results}

The fits are displayed in Fig.~\ref{fig:fuor-sed} and their fundamental
parameters in Table~\ref{tab:param-fit}.  For comparison's sake a standard
viscous disc model and a black-body backwarmed disc, that is with no
superheated surface, as in \citet{Kenyon91}, are shown along with the two-layer
model fit.  It appears that the black-body model gives a slightly higher
far-infrared excess and a smaller mid-infrared excess than the present model
and does not produce the 10\,{\micron} silicate feature. 

\subsubsection{FU~Ori}

Up to $10\,\micron$ the SED of FU~Ori can be modelled by a standard
viscous disc (Fig.~\ref{fig:fuor-sed}, dotted line of the top-left panel).  The
accretion rate  $\Mdot\Mstar = \sci{3.2}{-5}\,\Msun^2/\yr$ corresponds to
typical values in the literature \citep[$\approx 3/\correct{\cos i}\,\Msun^2/\yr$ where
$i$ is the inclination of the disc according to][]{Hartmann96,Malbet98}.
Nevertheless, the 5--12\,{\micron} SED is about 30\% weaker as expected from
the fit of 0.5--3\,{\micron} fluxes.

A black-body backwarmed disc \citep[see][]{Kenyon91} can explain
the far-IR excess but it either overestimates the 30--100\,{\micron} fluxes
(same figure, dashed line) or underestimates the 20--30\, and
100--200\,{\micron} SED (not shown).  If the backwarmed disc model comprises a
hot surface resulting from direct irradiation, then the SED can be convingly
fitted in the range 20--200\,{\micron} (solid line) with $H/r \approx 0.2$ at
10\,{\AU} from the central star.  This fit, however, overestimates the M and N
fluxes by a factor 2.5.  Interestingly, the model predicts a $10\,\micron$
silicate feature with the same intensity as in the observations, but with an
overestimated continuum.

That this centro-symmetric disc model does not accurately reproduce the
observations should not be taken too seriously as an argument against
backwarming, since the target appears to be a rather complicated object:
\citet{Malbet98} found by means of optical long-baseline interferometry that
the disc should present a hot spot located 15\,{\AU} from the star with a flux
ratio of 4--5 magnitudes in K; \citet{Wang04} also detected a companion at
200\,{\AU} (0.5'') with a similar flux ratio.

\subsubsection{V1057~Cyg}

V1057~Cyg is well described with an accretion disc featuring an
accretion rate $\Mstar\Mdot = \sci{2.0}{-5}\,\Msun^2/\yr$ \citep[in the range
predicted by][]{Kenyon88}, as shown in Fig.~\ref{fig:fuor-sed} (top-right panel,
dotted line). Backwarming is needed to account for the SED in the range
10--100\,{\micron} with a flaring $H/r \approx 0.35$ at 10\,{\AU} from the
star.  The black-body approach either overestimates the 20--50\,{\micron}
fluxes by a factor 3 (dashed line) or underestimates the 50--100\,{\micron} SED
by a similar factor (not shown).  The two-layer disc model fits the data well
in the range 1--100\,{\micron} (solid line).  It is worth noticing that the
model does produce a strong silicate feature while the flux in the N-band seems
to hint to the contrary --- it is at the level of the continuum predicted by the
model.

It is however worth noticing that this disc, though comparable to FU~Ori in
terms of accretion properties, presents a much larger flaring.  If, for some 
unknown reason, it features a lower viscosity, this could account for a larger
disc mass and a more flared structure.  On the other hand, it cannot be ruled out
that part of the far-IR excess is not to be imputed to backwarming and could
result from other phenomena, like an envelope or self-gravity-related energy
dissipation, in which case the flaring requirement is only an upper limit.

\subsubsection{V1515~Cyg}

V1515~Cyg presents an SED very similar to that of V1057~Cyg at
$\lambda \lesssim 20\,\micron$ but shifted down by a factor 5, which the larger
distance roughly explains.  It is therefore not surprising to find a similar
accretion rate $\Mstar\Mdot = \sci{1.3}{-5}\,\Msun^2/\yr$ with a standard disc
model (Fig.~\ref{fig:fuor-sed}, dotted line in the bottom-left panel).  Despite
of the lack of data at $\lambda \gtrsim 20\,\micron$, the 10--20\,{\micron}
fluxes require backwarming with a flaring $H/r \approx 0.35$ at 10\,{\AU} from
the star.  However, it is not possible to convincingly discriminate between the
black-body and the two-layer disc models.

\subsubsection{Z~CMa}

Z~CMa is a close (0.1'') binary \citep{Koresko89} YSO with FUor
characteristics.  The total flux can be described as emerging from an
intensively accreting FUor ($\Mdot \sim 10^{-3}\Msun/\yr$) for $\lambda
\lesssim 10\,\micron$, yet it fails at reproducing the almost flat far-IR SED.
A closer look at the SED of the individual components \citep[see][and
references therein]{Thiebaut95} hint that the primary is a cool YSO in its
primary phase of evolution (it peaks at 5\,\micron) while the secondary
features an FU~Ori-like SED.  Z~CMa~B is fairly well described with an accretion
disc $\Mstar\Mdot = \sci{1.4}{-5}\,\Msun/\yr$, but the lack of data at $\lambda
\gtrsim 5\,\micron$ rules out checking the backwarming hypothesis
(Fig.~\ref{fig:fuor-sed}, bottom-right panel).

\subsection{Reliability of the model}

This two-layer model proves that backwarming is a credible explanation for the
mid- and far-IR excess among a few FUors, without the need to invoke the
contribution of a circumstellar envelope.  It does not self-consistently
tackle the determination of the vertical structure, so I discuss here
the possible caveats linked to these simplifications.

\subsubsection{Simplification of the vertical structure}

Freeing the flaring parameter $H/h$ may seem hazardous, since its value
strongly affects the amount of irradiation caught by the disc and therefore the
strength of the IR excess \citep[ex.][for backwarmed discs]{Kenyon91}, so a
self-consistent determination of the irradiated surface by more elaborate
simulations \citep{DAlessio98,Dullemond02} may seem more able to corroborate or
invalidate the backwarming scenario.  Yet the latter simulations hide  a
flaring hypothesis, since they depend on the up-to-now poorly constrained mass
column in the disc; in other words, they do not help to rule out or support the
backwarming scenario better than an empirical approach, as far as the diagnosis
is based on the SED.  In the future, they should rely on better constrained
masses and provide a reliable diagnostic, when the Atacama Large Millimetre
Array (ALMA) probes the disc mass at the 10\,\AU-scale or when IR long-baseline
interferometry measures flaring-related asymmetries at the \AU-scale ---\,for
instance with AMBER on the Very Large Telescope Interferometer (VLTI). 

The two-layer approximation may also seem problematic, but \citet{Dullemond03}
showed that its SED predictions stand close to those of full transfer
simulations.  On the contrary, it is not fit for spectroscopic studies
\citep{Aikawa02} since spectral lines are very sensitive to the temperature and
density profiles; in particular, it cannot be used to issue a diagnostic using
the strength and shape of the silicate feature.

\subsubsection{Ignoring the disc's own gravity}

\begin{table}[t]
   \centering 
   \def\mc#1{\multicolumn{2}{c}{#1}}
   \def\z{\phantom{0}}
   \caption{%
      Disc mass in FUor models: column density $\Sigma$ at 1 and 10\,{\AU} from
      the star, cumulated disc mass $\Mdisk$ within 10 and 100\,\AU, and
      radius {\Rsg} at which the disc's vertical gravity field equals that of
      the star.  Masses are lower values and radii {\Rsg} upper values, 
      obtained with a high viscosity parameter $\alpha = 0.1$.
   }%
   \label{tab:mass-fit}
   \begin{tabular}{lccccc}
      \hline\hline
                & \mc{$\Sigma$ ($10^3\,\kg\,\meter^{-2}$)} 
                                         & \mc{{\Mdisk} (\Msun)} & {\Rsg} (\AU)\\
                & 1\,\AU     &  10\,\AU  & 10\,\AU & 100\,\AU    &      \\
      \hline
      FU~Ori    &     21     &    \z4    & 0.020& 0.54           & 29\\
      V1057~Cyg &     15     &    \z3    & 0.015& 0.32           & 50\\
      V1515~Cyg &     10     &    \z2    & 0.010& 0.24           & 61\\
      Z~CMa~B   &     11     &    \z2    & 0.012& 0.31           & 40\\
      \hline
   \end{tabular}
\end{table}

FUor disc models predict large masses, as shown in Table~\ref{tab:mass-fit}, so
that the gravity of the disc can overwhelm the stellar gravity in the outer,
massive parts of the disc \citep[self-gravitating discs, see][and references
therein]{Bertin99,Hure00}. These works single out three effects:
\begin{enumerate}
   \item \label{enum:sg-vert}
   An additional vertical gravitational field that flattens the
   outer parts of the disc and should prevent them from being illuminated
   by the central source of radiation \citep{Hure00}, which would include
   backwarming from inner disc regions.
   \item \label{enum:sg-horiz}
   An additional horizontal gravitational field so that the disc
   is no longer in Keplerian rotation.  This affects the viscous dissipation
   and the distribution of the matter. \citep{Bertin99}
   \item \label{enum:sg-heat}
   Possible instabilities in self-gravitating regions.
   The disc is then more likely to depart from a symmetrical geometry and
   presents an additional dissipation of energy, hence an enhanced
   IR excess.  \citet{Lodato01} proposed this phenomenon to explain the 
   far-IR excess among FUors.
\end{enumerate}
Phenomenon \ref{enum:sg-vert} is critical when determining the SED of a
backwarmed disc.  For each model fit, I computed an estimate of the radius
{\Rsg} at which the disc undergoes a transition toward self-gravity, namely
where the vertical gravity field produced by the disc at its surface equals
that of the central star.  As shown in Table~\ref{tab:mass-fit}, the three
successfully fit FUors should have $\Rsg \approx 50\,\AU$ with the hypothesis
of a high viscosity coefficient $\alpha = 0.1$.  In this case, self-gravity
should mostly affect the coolest regions and therefore the SED at $\lambda
\gtrsim 50\,\micron$.  Phenomena \ref{enum:sg-horiz},\ref{enum:sg-heat} mostly
affect viscous dissipation and can account for the far-IR SED for $\alpha \sim
0.01$ \citep{Lodato01}.  It is not impossible that the nature of the IR excess
should be imputed to contributions of both backwarming and self-gravity.

It should be emphasised that no study of self-gravity included the influence of
irradiation. The latter possesses a conterweight effect:  by heating the outer
parts of the disc, it increases the flaring and ---\,at least within the
$\alpha$ viscosity prescription\,--- decreases the amount of matter needed to
sustain a given accretion rate.  Furthermore, it is not clear whether FU~Ori
discs are as massive as expected from standard steady-disc models: it might
well happen during the outbursts that only the innerer parts undergo a sudden
increase of the accretion rate, while the outer parts of the disc conserve
TTS-like properties \citep{Bell95}.

\section{Conclusion}

Using the viscous and irradiated disc model by \citet{Lachaume03b},
it is possible to account for the strength of the mid- and far-IR excess
among a few FUors using the viscous luminosity of the inner parts of
the disc as an irradiation source.  Yet the model makes several approximations
in the determination of the structure:  the temperature profile is assimilated
to two isothermal layers, a surface and an interior, which should not prove
critical in the SED diagnostic; the flaring of the disc has been left
as a free parameter, but it proves no less relevant than full numerical
simulations, that assume the unknown amount of material in accretion
discs at a scale $\lesssim 100\,AU$;  more critical is my leaving aside
the self-gravity of the disc, which could ``unflare'' the disc and prevent
irradiation, but the simulation of self-gravity in the presence of irradiation
has not yet been performed, so its influence still remains speculation.

This work can be seen as a feasibility study for a future vertical structure
simulation of irradiated discs that I am developping using the radiative
transfer formalism presented in \citet{Malbet01}.  Such a simulation is needed
to issue a spectral diagnostic of irradiation (eg. silicate feature in
emission), that forthcoming IR long-baseline insterferometers will ease with
their AU-scale resolution for the closest FUors.  In particular, MIDI on the
VLTI will be able to measure the silicate feature while disentangling the
contribution of an envelope from that from an irradiated disc, which an SED
diagnostic cannot do \citep{Vinkovic03}.  One also expects to obtain
constraints on the flaring with IR closure phases (AMBER on the VLTI) and on
the disc mass with ALMA; their determination is not model-independent and also
requires a reliable model.

\begin{acknowledgement}
   This work has made use of NASA's Astrophysics Data System Bibliographic
   Services and of CDS's Vizier Catalogue Database.  Computations and graphics
   have been done with free software, in particular Yorick by D.~Munro.
   I also wish to thank C.~P.~Dullemond for helpful comments that
   improved the quality of the paper.  Language corrections have been
   suggested by K.~Smith.
\end{acknowledgement}

\bibliography{biblio}

\end{document}